\documentstyle[aps,prb]{revtex}
\begin{document}

\title{Transverse spin dynamics in a spin-polarized Fermi liquid}
\author{V.P.Mineev} \address{Commissariat a l'Energie Atomique, DSM,
Departement de Recherche Fondamentale sur la Matiere Condensee, SPSMS,
38054 Grenoble, France } \date{ Submitted 1 September 2003}
\maketitle
\begin{abstract}
The linear equations for transverse spin dynamics in a weakly polarized
degenerate Fermi liquid with arbitrary relationship between temperature $T$ 
and
polarization $\gamma H$ are derived from Landau-Silin phenomenological
kinetic equation with general form of two-particle collision integral. 
The temperature and polarization dependence of the spin current
relaxation time is established.  It is found in particular that at finite
polarization transverse spin wave damping has a finite value at $T=0$.

The analogy between temperature dependences of spin waves
attenuation and ultrasound absorption in degenerate Fermi liquid at
arbitrary temperature is presented.

We also discuss spin-polarized Fermi liquid in the general context of the
Fermi-liquid theory and compare it with
"Fermi liquid" with spontaneous magnetization.

\end{abstract}
\bigskip
PACS numbers:71.10.Ay, 67.65.+z, 67.55.Hc, 67.55.Jd

\bigskip

\section{Introduction}

The relaxation properties in a degenerate Fermi liquid are determined by the
collisions of quasiparticles.  Due to the Pauli exclusion principle 
only the quasiparticles near the Fermi surface in a layer with thickness
of the order of temperature are effectively exchanged by energy and
momentum.  Consequently, the relaxation time is proportional to $T^{-2}¥$
that leads to the temperature dependences of kinetic coefficients
\cite {1,2} of viscosity $\eta\propto T^{-2}¥$ and thermalconductivity
$\kappa\propto T^{-1}¥$.  The longitudinal spin-diffusion coefficient
$D_{\parallel}$ determining the spin current in presence of a gradient
of the absolute value of magnetization has been found \cite{3}
proportional to $¥\propto T^{-2}¥$.  So, kinetic coefficients diverge
when temperature tends to zero.  A similar situation takes place in a
spin-polarized Fermi liquid, where there are two Fermi distributions
for spin-up and spin-down particles with different Fermi momenta
$p_{F}¥^{\uparrow}¥$ and $p_{F}¥^{\downarrow}¥$.  So long we deal with
relaxation processes being determined by the collisions of
quasiparticles from thermal vicinity of one of the Fermi surfaces
(scattering of two quasiparticles with spins up (down)) or scattering
of quasiparticles from thermal vicinities of two different Fermi
surfaces (scattering of spin up and spin down quasiparticles ), \cite
{4,5,6}, the relaxation time and kinetic coefficients of viscosity and
longitudinal spin diffusion are proportional to $\propto T^{-2}¥$ and
the thermalconductivity to \linebreak  $\propto T^{-1}¥$.

Another type of relaxation process characterizes 
the spin current due to gradient of direction of magnetization or
so-called transverse
spin diffusion \cite{7}.  Indeed at $T=0$ all the states with
momenta below the smaller of two Fermi-surfaces (say for
$p<p_{F}¥^{\downarrow}¥$) and with plus one-half or minus one-half
projections of the spin to {\it an arbitrary oriented quantization
axis} are completely occupied: $W^{\uparrow}¥_{{\bf
p}}¥=W^{\downarrow}¥_{{\bf p}}¥=1$.  Hence the inhomogeneous rotation
of magnetization does not change the equilibrium state of
quasiparticles with momenta $p<p_{F}¥^{\downarrow}¥$.  But for the
Fermi particles with spin up and momenta in between two Fermi surfaces
$p_{F}¥^{\downarrow}¥<p<p_{F}¥^{\uparrow}¥$ the probability to have
spin-up projection to the rotated quantization axis, determined by
equilibrium direction of magnetization, is deviated from unity.  Hence
the relaxation process should involve all such particles even at
$T=0$.  In Fermi liquid this independent particle picture is changed
due to interaction creating a dissipationless inhomogeneous
magnetization rotation.  However, in the presence of finite
polarization a dissipative transverse diffusion motion is also
present.  Corresponding relaxation time does not diverge at zero
temperature \cite{7,8,9,10,11} and transverse spin waves attenuate at
$T=0$.

The calculations of transverse spin-diffusion coefficient $D_{\perp}¥$
have been done in dilute degenerate Fermi gas with arbitrary
polarization at $T=0$ in the papers by W.Jeon and W.Mullin \cite{7},
A.Meyerovich and K.Musaelian \cite{8}, and at $T\ne 0$ in the article
\cite{9}.  A derivation and an exact solution of the kinetic equation
in the s-wave scattering approximation for dilute degenerate Fermi gas
with arbitrary polarization at $T=0$ and for a small polarization $\mu
H \ll \varepsilon_{F}¥$ at $T\ne 0$ have been obtained also in the
papers by D.Golosov and A.Ruckenstein \cite{10}.  For the treatment of
this problem in a Fermi liquid the Matthiessen-type rule arguments
(sum of temperature-driven and polarization-driven scattering rates)
and simple relaxation-time approximation for the collision integral
have been used \cite{11}.  Thus the zero temperature attenuation of
the transverse spin waves has been established.

This conclusion has been contested by I.Fomin  \cite{12},  
who has proposed dissipationless spin-wave dispersion relation
\begin{equation}
\omega=\omega_{L}¥ +\chi k^{2}¥,
\label{eA}
\end{equation}
where $\chi$ is a coefficient of proportionality between spin current and
the chiral spin velocity arising at an inhomogeneous rotation of spin
space.  Fomin has not presented a calculation of this value.  He just
has established the spin-wave dispersion law in the assumptions that in
a polarized Fermi liquid at $T=0$ all the spin current is the chiral
current and it can be derived as response to the generalized gauge
transformation or inhomogeneous spin space rotation like it has been done for
superfluid $^{3}¥He$ \cite {13,14}.

Certainly, if in the process of such a type of derivation we shall ignore
the quasiparticles finite scattering rate, one can obtain
the dissipationless spin current originating from Fermi-liquid
interaction.  However, besides the reactive part the total spin
current calculated in presence of collisions includes dissipative or
spin-diffusion part resulting in imaginary part of dispersion law for
the transverse spin waves.  At the microscopic level the collisions
treatment is equivalent to derivation of kinetic equation that was
done for the case of spin-polarized diluted Fermi gas \cite
{7,8,9,10}.

Being addressed to the same problem in polarized Fermi-liquid, we need
the kinetic equation.  The derivation of it for a strongly interacting
Fermi liquid is unreal problem.  However, for a weakly polarized Fermi
liquid it seems natural to work on the basis of semi-phenomenological
Silin-type kinetic equation \cite{15}.  Assuming its validity in the
present paper we reexamine the derivation of transverse spin dynamics in
weakly polarized Fermi liquid for arbitrary relationship between
temperature $T$ and polarization $\gamma H$.  At small space and time
variations of transversal part of vectorial quasiparticle distribution
function we shall obtain Leggett-type \cite{16} equations for spin and
spin current densities.  Then from the general form of two-particle
collision integral similar to that was derived in \cite{17,18} we
deduce the spin current relaxation term.  The latter is essentially
simplified when both the temperature $T$ and the polarization $\gamma
H$ are much smaller than Fermi energy $\varepsilon_{F}¥$ and the
momenta of all excitations are confined to lie in the vicinity of both
Fermi surfaces and therefore one may decouple the angular and energy
variables in the collision integral in the manner first introduced by
Abrikosov and Khalatnikov \cite{2}.

We confirm the results of the papers \cite{7,8,9,10} where the same problem 
were
treated for the dilute Fermi gas.  It is found in particular that
at finite polarization spin-wave damping has a finite value at $T=0$.
More precisely, at low temperatures it proves to be proportional to the
number of collisions between quasiparticles 
\begin{equation}
\frac{1}{\tau}~ \propto~ ((\gamma H)^{2}¥+(2\pi T)^{2}¥).
\label{eB}
\end{equation}
This corresponds to the law of zero sound attenuation 
\cite{19}
\begin{equation}
\gamma ~ \propto~ (\omega ^{2}¥+(2\pi T)^{2}¥),
\label{eC}
\end{equation}
which is also determined by the number of collisions between quasiparticles. 
One can find the results of recent measurements of low-temperature zero sound
attenuation and surway of previous experimental works on this subject
in the paper \cite{20}.

In general, all regimes of the temperature behavior of the spin-wave
absorption in the degenerate Fermi liquid can be juxtaposed with
correspondent regimes in the absorption of ultrasound.  We shall
consider this analogy in the conclusion.

At the end we shall discuss spin-polarized Fermi liquid in the general context
of the Fermi-liquid theory and compare it with an imaginary
ferromagnetic Fermi liquid or liquid with spontaneous magnetization.

\section {Spin dynamics equations}

The quasiparticle distribution function as well as quasiparticle
energy are given by $2\times2$ matrix in spin space,
\begin{equation}
\hat n_{\bf k}¥({\bf r},t)=n_{\bf k}¥({\bf r},t)\hat I +
\mbox{\boldmath $\sigma $}_{\bf k}¥({\bf r},t) \mbox{\boldmath
$\hat \sigma $},
\label{e1}
\end{equation}
\begin{equation}
\hat \varepsilon_{\bf k}¥({\bf r},t)=
\varepsilon_{\bf k}¥({\bf r},t)\hat I +
{\bf h}_{\bf k}¥({\bf r},t) \mbox{\boldmath$\hat \sigma$}.
\label{e2}
\end{equation}
Here $\mbox{\boldmath$\hat\sigma$}$= $({\hat\sigma}_{x}¥,
{\hat\sigma}_{y}¥,{\hat\sigma}_{z}¥)$ are Pauli matrices.  The scalar and
vector parts of matrix distribution function obey the coupled kinetic
equations \cite{15}.  As it was pointed out by Leggett \cite{16} in
the case of small polarizations, the equation for the scalar part of the
distribution function $n_{\bf k}¥({\bf r},t)$ 
decouples from the equation for the vector part of distribution function
$\mbox{\boldmath $\sigma $}_{\bf k}¥({\bf r},t)$ and we may put
$n_{\bf k}¥$ equal to its equilibrium value, namely, usual Fermi
function.  At that the equation for 
$\mbox{\boldmath $\sigma $}_{\bf k}¥({\bf r},t)$ still shall be
nonlinear: as long as the polarization is small by its absolute value one can
consider the arbitrary large variations of the direction of
magnetization.  On the other hand, the similar decoupling of the
equation for the scalar part of the distribution function from the equation for
the vector part of distribution function including the collision
integrals (see below) takes place at arbitrary polarizations as long
as we consider the small deviations of the magnetization direction from
its equilibrium direction.  We shall be interested in the latter case.

In general, the equation for the $\mbox{\boldmath $\sigma $}_{\bf
k}¥({\bf r},t)$ has the form
\begin{eqnarray}
\frac{\partial \mbox{\boldmath$\sigma$}_{\bf k}¥}{\partial t}& +&
\frac{\partial\varepsilon_{\bf k}¥}
{\partial k_{i}¥}
\frac{\partial \mbox{\boldmath$\sigma$}_{\bf k}¥}{\partial x_{i}¥}-
\frac{\partial\varepsilon_{\bf k}¥}
{\partial x_{i}¥} \frac{\partial \mbox{\boldmath$\sigma$}_{\bf
k}¥}{\partial k_{i}¥}+
\frac{\partial {\bf h}_{\bf k}¥} {\partial k_{i}¥} \frac{\partial
n_{\bf k}¥}{\partial x_{i}¥}- \frac{\partial {\bf h}_{\bf k}¥}
{\partial x_{i}¥} \frac{\partial n_{\bf k}¥}{\partial k_{i}¥}\nonumber \\
&-&2({\bf h}_{\bf k}¥\times \mbox{\boldmath$\sigma$}_{\bf k}¥) 
=\left (\frac{\partial \mbox{\boldmath$\sigma$}_{\bf k}¥}{\partial t}
\right )_{coll}¥.
\label{e3}
\end{eqnarray}
We divide all matrices in equilibrium and nonequilibrium parts,
\begin{equation}
\hat n_{\bf k}¥=\hat n_{\bf k}¥^{0}¥+ \delta \hat n_{\bf k}¥ ,
\label{e4}
\end{equation}
\begin{equation}
\hat \varepsilon_{\bf k}¥= \hat \varepsilon_{\bf k}¥^{0}¥ +
\delta \hat \varepsilon_{\bf k}¥,
\label{e5}
\end{equation}
where
\begin{equation}
\hat n_{\bf k}¥^{0}¥=\frac{1}{2}(n_{0}¥^{+}¥+n_{0}¥^{-}¥)\hat I+
\frac{1}{2}(n_{0}¥^{+}¥-n_{0}¥^{-}¥)( \mbox{\boldmath$\hat\sigma$}
\hat {\bf m})
\label{e6}
\end{equation}
is the equilibrium distribution function of polarized Fermi liquid
and
\begin{equation}
\hat \varepsilon_{\bf k}¥^{0}¥ =\varepsilon_{\bf k}¥\hat I
-\frac{1}{2}\gamma({\bf B}\mbox{\boldmath$\hat\sigma$}) 
\label{e7}
\end{equation}
is the equilibrium quasiparticle energy.
Here there are two Fermi distribution
functions
\begin{equation}
n_{0}¥^{\pm}¥(\varepsilon_{\bf k}¥)=
n_{0}¥(\varepsilon_{\bf k}¥\mp \frac{\gamma H}{2})=
\frac{1}{\exp\left(\frac {\varepsilon_{\bf k}¥\mp \frac{\gamma
H}{2} -\mu }{T}\right ) +1}
\label{e8}
\end{equation}    
shifted on the value of polarization $\gamma H/2$, $\gamma$ is the gyromagnetic ratio, Planck constant $\hbar=1$
throughout the paper, the polarization direction is determined by the
unit vector ${\bf m}= {\bf H}/H$.

The "effective" magnetic field ${\bf H}$ is the field corresponding to
the magnetization created by the external magnetic field ${\bf
H}_{0}¥$ and by the pumping \cite{21}.  The pumped part in view of very
long time of longitudinal relaxation should be considered as
equilibrium part of magnetization.  The difference between ${\bf B}$
and ${\bf H}$ originates from the pumping that changes the
quasiparticle distribution functions but does not directly affect on the
energy of quasiparticles.  Field ${\bf B}$ consists of an external
magnetic field ${\bf H}_{0}¥$ and the Fermi-liquid molecular field.  To
define ${\bf B}$ we must consider the equilibrium distribution matrix
(\ref{e6}) and equilibrium energy matrix (\ref{e7}) as deviations from
the corresponding matrices for nonpolarized Fermi liquid,
\begin{equation}
\hat n_{\bf k}¥^{0}¥=n_{0}¥(\varepsilon_{\bf k}¥)\hat I+ \delta
\hat n_{\bf k}¥^{0}¥,
\label{e9}
\end{equation}
\begin{equation}
\hat \varepsilon_{\bf k}¥^{0}¥=\varepsilon_{{\bf k}}¥\hat I
-\frac{1}{2}\gamma({\bf B}\mbox{\boldmath$\hat\sigma$})=
\varepsilon_{{\bf k}}¥\hat I -\frac{1}{2}\gamma({\bf
H}_{0}¥\mbox{\boldmath$\hat\sigma$})+ \frac{1}{2}Sp'\int d\tau'f_{{\bf
k}{\bf k'}}¥^{\sigma \sigma'}¥ \delta \hat n_{\bf k'}¥^{0}¥,
\label{e10}
\end{equation}
where $d \tau=2d {\bf k}/(2\pi)^{3}¥$ and the Fermi-liquid matrix of
interaction is
\begin{equation}
f_{{\bf k}{\bf k'}}¥^{\sigma \sigma'}¥=
f_{{\bf k}{\bf k'}}¥^{s}¥\hat I\hat I'+
f_{{\bf k}{\bf k'}}¥^{a}¥
\mbox{\boldmath$\hat \sigma $}\mbox{\boldmath$\hat \sigma' $} +
f_{{\bf k}{\bf k'}}¥^{b}¥\hat {\bf m}
(\mbox{\boldmath$\hat \sigma $}\hat I'+
\hat I\mbox{\boldmath$\hat \sigma' $})+
f_{{\bf k}{\bf k'}}¥^{c}¥
(\hat {\bf m}\mbox{\boldmath$\hat \sigma $})
(\hat {\bf m}\mbox{\boldmath$\hat \sigma' $}). 
\label{e11}
\end{equation}    
These three equation together give an equation for ${\bf B}$ determination 
\begin{equation}
\gamma{\bf B}=\gamma{\bf H}_{0}¥- \hat{\bf m}\int d\tau' [(f_{{\bf k}{\bf
k'}}¥^{a}¥+f_{{\bf k}{\bf k'}}¥^{c}¥)
(n_{0}¥^{+}¥-n_{0}¥^{-}¥)+f_{{\bf k}{\bf k'}}¥^{b}¥
(n_{0}¥^{+}¥+n_{0}¥^{-}¥-2n_{o}¥)].
\label{e12}
\end{equation}
In the absence of a pumped magnetization the field, ${\bf B}={\bf H}$ and
(\ref{e12}) is just the self-consistency equation for the field ${\bf
H}$ determination as the function of an external field ${\bf H}_{0}¥$. 
When the part of magnetization is created by pumping, ${\bf H}$
presents an independent value and the total energy shift $\gamma ({\bf
B}\mbox{\boldmath$\hat \sigma $})/2$ is determined by means of two
fields: external ${\bf H}_{0}¥$ and "effective" ${\bf H}$.  We shall
assume that they are parallel each other.

For the finite polarization the vector ${\bf B}$ proves to be energy
dependent.  This reflects  the impossibility to formulate a Fermi-liquid theory
with finite polarization in terms of Landau Fermi-liquid parameters
which are just numbers characterizing the intensity of interaction of
quasiparticles near the Fermi surface.  The problem is not resolved
even by introduction of Fermi-liquid parameters separately for each
Fermi sphere with Fermi momenta $p_{F}¥^{\downarrow}¥$ and
$p_{F}¥^{\uparrow}¥$.  To avoid this complexity we shall limit ourselves by
the case of small polarization assuming independence of the functions
$f_{{\bf k}{\bf k'}}¥$ of energy.  Then it is clear that ${\bf B}$ is
energy independent and determined just by zeroth-order term in
expansion of the functions $f_{{\bf k}{\bf k'}}¥^{i}¥$ on spherical
harmonics.

We shall discuss the only
perpendicular deviations from the initial equilibrium state, 
\begin{equation}
\delta \hat n_{\bf k}¥=
\delta\mbox{\boldmath $\sigma $}_{\bf k}¥({\bf r},t) \mbox{\boldmath
$\hat \sigma $},  ~~~(\hat {\bf m}\delta\mbox{\boldmath $\sigma $}_{\bf k}¥)=0.
\label{e13}
\end{equation}    
Then the energy deviation matrix  has the form
\begin{equation}
\delta\hat \varepsilon_{\bf k}¥ =\delta{\bf h}_{\bf k}¥
\mbox{\boldmath$\hat \sigma$}, ~~~
\delta{\bf h}_{\bf k}¥=
\int d\tau'f_{{\bf k}{\bf k'}}¥^{a}¥ \delta\mbox{\boldmath
$\sigma $}_{\bf k'}¥
\label{e14}
\end{equation}
and the kinetic equation (\ref{e3}) can be rewritten as
\begin{equation}
\frac{\partial \delta \mbox{\boldmath$\sigma$}_{\bf k}¥}{\partial t} +
\frac{\partial\varepsilon_{\bf k}¥^{0}¥} {\partial
k_{i}¥}
\frac{\partial \delta
\mbox{\boldmath$\sigma$}_{\bf k}¥}{\partial x_{i}¥}- \frac{1}{2}
\frac{\partial(n_{0}¥^{+}¥+n_{0}¥^{-}¥)} {\partial
k_{i}¥}\frac{\partial \delta{\bf h}_{\bf k}¥}{\partial x_{i}¥}
-2\left [\left(-\frac{\gamma {\bf B}}{2}+\delta{\bf h}_{\bf k}¥\right)\times
\left(\frac{1}{2}(n_{0}¥^{+}¥-n_{0}¥^{-}¥)\hat {\bf m}+\delta
\mbox{\boldmath$\sigma$}_{\bf k}¥ \right)\right ] =\left (\frac{\partial
\mbox{\boldmath$\sigma$}_{\bf k}¥}{\partial t} \right )_{coll}¥.
\label{e15}
\end{equation}

We deal with linear in  $\delta\mbox{\boldmath $\sigma $}_{\bf k}¥$
equation with coefficients independent of space and time variables. 
In the lowest order on polarization, these coefficients are expressed
through Fermi-liquid parameters for nonpolarized Fermi liquid which
are introduced as usual by 
\begin{equation}
    f_{{\bf k}{\bf k'}}¥^{a}¥=
N_{0}¥^{-1}¥\sum_{l}¥F_{l}¥^{a}¥P_{l}¥(\hat{\bf k},\hat{\bf k'}),
\label{eD}
\end{equation}
where $N_{0}¥=m^{*}¥k_{F}¥/\pi^{2}¥$ is the density of states.  As
mentioned, for large polarizations the coefficients in the kinetic
equation are not well determined, although the structure of the
equation looks similar.  So, for simplicity we limit ourselves by the
treatment of the left hand side of the equation in the lowest order on
polarization.  One can neglect in this case $F^{b}¥_{l}¥$ and
$F^{c}¥_{l}¥$ .  Then following Leggett \cite{16} one may rewrite
equation (\ref{e15}) as two equations for the first two harmonics
(magnetization density and spin current density) of the distribution
function
\begin{equation}
{\bf M}({\bf r},t)=
\frac{1}{2}\int d \tau \delta \mbox{\boldmath$\sigma$}_{\bf k}¥, 
\label{e16}
\end{equation}
\begin{equation}
{\bf J}_{i}¥({\bf r},t)=\frac{1}{2}\int d \tau \left [v_{Fi}¥\delta
\mbox{\boldmath$\sigma$}_{\bf k}¥-\frac{\partial n_{0}¥}{\partial
k_{i}¥} \delta{\bf h}_{\bf k}¥\right ]= \frac{1}{2}( 1+\frac{F_{1}¥^{a}¥}{3} )
\int d \tau v_{Fi}¥\delta \mbox{\boldmath$\sigma$}_{\bf k}¥.
\label{e17}
\end{equation}
They are
\begin{equation}
\frac{\partial {\bf M}}{\partial t} +\frac{\partial 
{\bf J}_{i}¥}{\partial x_{i}¥} -{\bf M}\times \gamma {\bf H}_{0}¥=0,
\label{e18}
\end{equation}
\begin{eqnarray}
\frac{\partial {\bf J}_{i}¥}{\partial t} &+&
\frac{1}{3}v_{F}¥^{2}¥(1+F_{0}¥^{a}¥)(1+\frac{F_{1}¥^{a}¥}{3})
\frac{\partial{\bf M}}{\partial x_{i}¥} - {\bf J}_{i}¥\times\gamma{\bf
H}_{0}¥ +\frac{4}{N_{0}¥}(F_{0}¥^{a}¥-\frac{F_{1}¥^{a}¥}{3})({\bf J}_{i}¥
\times{\bf M}^{\parallel}¥)\nonumber \\
&=&\frac{1}{2}(1+\frac{F_{1}¥^{a}¥}{3})\int d \tau v_{Fi}¥
\left (\frac{\partial
\mbox{\boldmath$\sigma$}_{\bf k}¥}{\partial t} \right )_{coll}¥.
\label{e19}
\end{eqnarray}
Here
\begin{equation}
{\bf M}^{\parallel}¥=
\frac{\gamma N_{0}¥}{4}{\bf H}
\label{e20}
\end{equation}

So far we developed the theory for the homogeneous external field. 
In this case one can excite transverse spin waves in an infinite medium.  The
situation has been described in Appendix A in the paper \cite{22}. 
However, the theory is valid also when the magnetic field is coordinate
dependent by its absolute value and includes a small but in general
fast in time supplementary rf part directed in a perpendicular
direction:
\begin{equation}
{\bf H}_{0}¥({\bf r}, t)={\bf H}_{0}({\bf r})+ {\bf h}({\bf r}, t),
~~~{\bf H}_{0}({\bf r})=\hat z (H_{0}+\delta H_{0}({\bf r})), ~~~~(\hat z{\bf
h}({\bf r}, t))=0
\label{e21}
\end{equation}

This situation is typical for the spin-waves experiments (see, for
instance \cite{23}).  In this case we must introduce the additional longitudinal
deviations in the distribution function and the energy of
quasiparticles.  So, equations (\ref{e13}) and (\ref{e14}) are modified
as follows:
\begin{equation}
\delta \hat n_{\bf k}¥=
\delta\mbox{\boldmath $\sigma $}_{\bf k}¥({\bf r},t) \mbox{\boldmath
$\hat \sigma $}+\delta\sigma^{\parallel}¥_{\bf k}¥ ({\bf r},t) \hat
\sigma_{z}¥, ~~~(\hat z\delta\mbox{\boldmath $\sigma $}_{\bf k}¥)=0,
\label{e22}
\end{equation}
\begin{equation}
\delta\hat \varepsilon_{\bf k}¥ =-\frac{1}{2}\gamma (\hat z\delta
H_{0}({\bf r})+ {\bf h}({\bf r},
t))\mbox{\boldmath$\hat\sigma$}+\delta{\bf h}_{\bf k}¥
\mbox{\boldmath$\hat \sigma$}+\int d\tau'f_{{\bf k}{\bf k'}}¥^{a}¥
\delta\sigma^{\parallel}¥_{\bf k'}¥ \hat \sigma_{z}¥,~~~ \delta{\bf
h}_{\bf k}¥= \int d\tau'f_{{\bf k}{\bf k'}}¥^{a}¥
\delta\mbox{\boldmath $\sigma $}_{\bf k'}¥.
\label{e23}
\end{equation}
In linear approximation to the perpendicular to $\hat z$ deviations the
equations for the perpendicular and the parallel parts of the
distribution function including the collision integral are
independent.  Thus for the transversal part we return back to the
slightly modified system of equations (\ref{e18}), (\ref{e19}):
\begin{equation}
\frac{\partial {\bf M}}{\partial t} +\frac{\partial 
{\bf J}_{i}¥}{\partial x_{i}¥} -{\bf M}\times \gamma {\bf H}_{0}¥({\bf r})
-({\bf M}^{\parallel}¥+\delta{\bf M}^{\parallel}¥({\bf r}))\times \gamma
{\bf h}({\bf r}, t) =0,
\label{e24}
\end{equation}
\begin{eqnarray}
\frac{\partial {\bf J}_{i}¥}{\partial t} &+&
\frac{1}{3}v_{F}¥^{2}¥(1+F_{0}¥^{a}¥)(1+\frac{F_{1}¥^{a}¥}{3})
\frac{\partial{\bf M}}{\partial x_{i}¥} - {\bf J}_{i}¥\times\gamma{\bf
H}_{0}¥({\bf r}) +
\frac{4}{N_{0}¥}(F_{0}¥^{a}¥-\frac{F_{1}¥^{a}¥}{3})({\bf J}_{i}¥
\times {\bf
M}^{\parallel}¥)\nonumber \\
&=&\frac{1}{2}(1+\frac{F_{1}¥^{a}¥}{3})\int d \tau
v_{Fi}¥ \left (\frac{\partial \mbox{\boldmath$\sigma$}_{\bf
k}¥}{\partial t} \right )_{coll}¥,
\label{e25}
\end{eqnarray}
where $\delta{\bf M}^{\parallel}¥({\bf r})$ is a change of spin density due to 
$\delta H({\bf r})$.

\section {Collision integral treatment}

The collision integral for the vectorial part of distribution function
is determined through the general collision integral for the matrix
distribution function as follows
\begin{equation}
\left (\frac{\partial
\mbox{\boldmath$\sigma$}_{\bf k}¥}{\partial t} \right )_{coll}¥=
\frac{1}{2}Sp\mbox{\boldmath$\hat \sigma $}\hat I_{coll}¥,
\label{e26}
\end{equation}
\begin{equation}
I^{\alpha \beta}¥_{coll}¥=-\frac{1}{4}\int d \tau'_{1}¥d\tau_{2}¥d {\bf k'}_{2}¥
\delta(\varepsilon_{1}¥+\varepsilon_{2}¥-\varepsilon'_{1}¥-
\varepsilon'_{2}¥) \delta({\bf k}+{\bf k'}_{1}¥-{\bf k}_{2}¥-{\bf
k'}_{2}¥) F^{\alpha \beta}¥({\bf k},{\bf k'}_{1}¥,{\bf k}_{2}¥,{\bf
k'}_{2}¥).
\label{e27}
\end{equation}
Here in absence of relativistic interactions $\varepsilon_{1}¥=
\varepsilon_{{\bf k}}¥$, $\varepsilon_{2}¥=
\varepsilon_{{\bf k_{2}¥}}¥$, etc. For only two-particle collisions
the function $F^{\alpha \beta}¥$ must contain the products of the
matrix distribution functions of the general form
$$
(\hat n_{{\bf k}}¥)^{\lambda_{1}¥ \lambda_{2}¥}¥ (\hat I-\hat
n_{{\bf k'}_{1}¥}¥)^{\lambda_{3}¥ \lambda_{4}¥} (\hat n_{{\bf
k}_{2}¥}¥)^{\lambda_{5}¥ \lambda_{6}¥}¥ (\hat I-\hat n_{{\bf
k'}_{2}¥}¥)^{\lambda_{7}¥ \lambda_{8}¥}
$$
multiplied on some tensor function depending of all momenta and the
spin indices.  However, in absence of relativistic interactions in such
the products the only matrix products of two types are possible:
$$
\frac{1}{2}\left \{[\hat n_{{\bf k}}¥(\hat I - \hat n_{{\bf
k'}_{1}¥}¥)]^{\alpha\beta}¥ 
+ [(\hat I - \hat n_{{\bf k'}_{1}¥}¥)\hat n_{{\bf
k}}¥]^{\alpha\beta}¥ \right \}Sp[\hat n_{{\bf k}_{2}¥}¥(\hat I-\hat n_{{\bf
k'}_{2}¥}¥)] $$
and $$\frac{1}{2}\left \{[\hat n_{{\bf k}}¥(\hat I - \hat n_{{\bf k'}_{1}¥}¥) \hat n_{{\bf
k}_{2}¥}¥(\hat I-\hat n_{{\bf k'}_{2}¥}¥)]^{\alpha\beta}¥+[(\hat I -
\hat n_{{\bf k'}_{1}¥}¥) \hat n_{{\bf k}_{2}¥}¥(\hat I-\hat n_{{\bf
k'}_{2}¥}¥)\hat n_{{\bf k}}¥]^{\alpha\beta}¥ \right \} $$
As usual, in quantum mechanics one must take the symmetrized (Hermitian)
products of operators and corresponding matrices .  Adding to this
expressions describing the scattering processes "going out" of initial
state the corresponding expressions for processes for "going in"
initial state, we can write the general form for the function $\hat F$
determining the scattering integral for the binary collisions
\begin{eqnarray}
F^{\alpha \beta}¥&(&{\bf k},{\bf k'}_{1}¥,{\bf k}_{2}¥,{\bf k'}_{2}¥)
\nonumber \\
&=&
\frac{1}{2}W_{1}¥({\bf k},{\bf k'}_{1}¥,{\bf k}_{2}¥,{\bf k'}_{2}¥)
\left (
\{[\hat n_{{\bf k}}¥(\hat I - \hat n_{{\bf k'}_{1}¥}¥)]^{\alpha\beta}¥
+ [(\hat I -
\hat n_{{\bf k'}_{1}¥}¥)\hat n_{{\bf k}}¥]^{\alpha\beta}¥\} Sp[\hat
n_{{\bf k}_{2}¥}¥(\hat I-\hat n_{{\bf k'}_{2}¥}¥]\right.  \nonumber \\
&-& 
\left.\{[(\hat I-\hat n_{{\bf k}}¥)\hat n_{{\bf k'}_{1}¥}¥]^{\alpha\beta}¥
+ [\hat
n_{{\bf k'}_{1}¥}¥(\hat I-\hat n_{{\bf k}}¥)]^{\alpha\beta}¥ \}Sp[(\hat
I -\hat n_{{\bf k}_{2}¥}¥)\hat n_{{\bf k'}_{2}¥}¥]\right)\nonumber \\
&+&
\frac{1}{2}W_{2}¥({\bf k},{\bf k'}_{1}¥,{\bf k}_{2}¥,{\bf k'}_{2}¥) \left(
[\hat n_{{\bf k}}¥(\hat I - \hat n_{{\bf k'}_{1}¥}¥) \hat n_{{\bf
k}_{2}¥}¥(\hat I-\hat n_{{\bf k'}_{2}¥}¥)]^{\alpha\beta}¥+[(\hat I -
\hat n_{{\bf k'}_{1}¥}¥) \hat n_{{\bf k}_{2}¥}¥(\hat I-\hat n_{{\bf
k'}_{2}¥}¥)\hat n_{{\bf k}}¥]^{\alpha\beta}\right.\nonumber \\
&-& 
\left.[(\hat I-\hat n_{{\bf k}}¥) \hat n_{{\bf k'}_{1}¥}¥(\hat I- \hat
n_{{\bf k}_{2}¥}¥)\hat n_{{\bf k'}_{2}¥}¥]^{\alpha\beta}¥-[\hat
n_{{\bf k'}_{1}¥}¥(\hat I- \hat n_{{\bf k}_{2}¥}¥)\hat n_{{\bf
k'}_{2}¥}¥ (\hat I-\hat n_{{\bf k}}¥)]^{\alpha\beta}\right)¥.
\label{e28}
\end{eqnarray}
Due to the total quasiparticle
density $Sp\int d {\bf k} \hat I_{coll}¥$ and the total quasiparticle
spin density $Sp\mbox{\boldmath$\hat \sigma $}\int d {\bf k} \hat
I_{coll}¥/2$ 
conservation, the functions $W_{1}¥$ and $W_{2}¥$ obey the following
conditions $W_{1}¥({\bf k}, {\bf k'}_{1}¥,{\bf k}_{2}¥,{\bf k'}_{2}¥)=
W_{1}¥({\bf k'}_{1}¥,{\bf k},{\bf k'}_{2}¥,{\bf k}_{2}¥)$ and
$W_{2}¥({\bf k},{\bf k'}_{1}¥,{\bf k}_{2}¥,{\bf k'}_{2}¥)= W_{1}¥({\bf
k'}_{1}¥,{\bf k}_{2}¥,{\bf k'}_{2}¥,{\bf k})$.  The two-particle
collision integral for the matrix distribution function determined by
(\ref{e27}), (\ref{e28}) corresponds to collision integral derived in
Born approximation by V.Silin \cite{17} (see also \cite{18}).  When
all the distribution function matrices are diagonal, the collision
integral (\ref{e22}), (\ref{e25}) reduces to the diagonal form of 
two-particle collision integral \cite{24} with
$W_{1}¥=2W_{\uparrow\downarrow}¥$ and
$W_{1}¥+W_{2}¥=W_{\uparrow\uparrow}¥$.

For the case of low temperature $T\ll\varepsilon_{F}¥$ and small
polarization $\gamma H \ll\varepsilon_{F}¥$ when all the quasiparticle
momenta of scattering particles lie near the Fermi surface of
nonpolarized Fermi liquid one can suppose as in the paper \cite{2}
that functions $W_{1}¥$ and $W_{2}¥$ depend only on the angle $\theta$
between $\bf k$ and ${\bf k}_{2}¥$ and on the angle $\phi$ between the
planes $({\bf k},{\bf k}_{2}¥)$ and $({\bf k'}_{1}¥,{\bf k'}_{2}¥)$. 
Now we must take (\ref{e28}) in the linear approximation on deviations
$\delta \hat n$.  We deal only with the terms containing deviations
$\delta \hat n_{\bf k}¥= \delta\mbox{\boldmath $\sigma $}_{\bf
k}¥({\bf r},t) \mbox{\boldmath $\hat \sigma $}$ and do not consider 
the terms containing deviations $\delta \hat n_{\bf k'_{1}¥}¥$ etc because
after all integrations in the collision integral (\ref {e27}) they are
independent of the ${\bf k}$ direction ( equilibrium distribution
matrix $\hat n_{\bf k}¥^{0}¥$ is isotropic in the ${\bf k}$ space). 
Hence they disappear at final stage after integration in the right
hand side of (\ref{e25}).  We choose the local direction of the
quantization axis $\hat {\bf m}$ along $\hat z$ direction, such that
$(\hat z\delta\mbox{\boldmath $\sigma $}_{\bf k}¥)=0$ .  We can
perform the integration over ${\bf k'}_{2}¥$ in (\ref{e27})
eliminating delta function of momenta and also, following the procedure
of the article \cite{2} reproduced in review \cite{25} in somewhat
different manner, reexpress the integration over momentum space as
$$
d{\bf k_{2}¥}d{\bf k'_{1}¥}=\frac{(m^{*}¥)^{3}¥}{2\cos(\theta/2)}
d\varepsilon_{2}¥d\varepsilon'_{1}¥d\varepsilon'_{2}¥\sin\theta
d\theta d\phi d\phi_{2}¥
$$
So, the linear part of the collision
integral is
\begin{eqnarray}
\delta \hat I_{coll}¥=-&\delta &\hat n_{\bf
k}¥\frac{m^{*}¥^{3}¥}{2(2\pi)^{5}¥} \int
d\varepsilon_{2}¥d\varepsilon'_{1}¥d\varepsilon'_{2}¥
\delta(\varepsilon_{1}¥+\varepsilon_{2}¥-\varepsilon'_{1}¥-
\varepsilon'_{2}¥) \left\{\overline {W_{1}¥} \left\{
[1-n_{0}¥^{+}¥(\varepsilon'_{1}¥)\right.\right.  \nonumber\\
&+&\left.\left. 1-n_{0}¥^{-}¥(\varepsilon'_{1}¥)]
[n_{0}¥^{+}¥(\varepsilon_{2}¥)(1-n_{0}¥^{+}¥(\varepsilon'_{2}¥))+
n_{0}¥^{-}¥(\varepsilon_{2}¥)(1-n_{0}¥^{-}¥(\varepsilon'_{2}¥))]\right.\right.\nonumber\\
&+&
\left.\left.[ n_{0}¥^{+}¥(\varepsilon'_{1}¥)+
n_{0}¥^{-}¥(\varepsilon'_{1}¥)]
[(1-n_{0}¥^{+}¥(\varepsilon_{2}¥))n_{0}¥^{+}¥(\varepsilon'_{2}¥)+
(1-n_{0}¥^{-}¥(\varepsilon_{2}¥))n_{0}¥^{-}¥(\varepsilon'_{2}¥)]\right\}\right.\nonumber\\
&+&
\overline {W_{2}¥}
\left.\left\{ (1 -n_{0}¥^{+}¥(\varepsilon'_{1}¥))
n_{0}¥^{+}¥(\varepsilon_{2}¥)(1-n_{0}¥^{+}¥(\varepsilon'_{2}¥)) +(1
-n_{0}¥^{-}¥(\varepsilon'_{1}¥))
n_{0}¥^{-}¥(\varepsilon_{2}¥)(1-n_{0}¥^{-}¥(\varepsilon'_{2}¥))\right.\right.\nonumber\\
&+&
\left.\left.  n_{0}¥^{+}¥(\varepsilon'_{1}¥)
(1-n_{0}¥^{+}¥(\varepsilon_{2}¥))n_{0}¥^{+}¥(\varepsilon'_{2}¥) +
n_{0}¥^{-}¥(\varepsilon'_{1}¥)
(1-n_{0}¥^{-}¥(\varepsilon_{2}¥))n_{0}¥^{-}¥(\varepsilon'_{2}¥)\right\}\right\},
\label{e29}
\end{eqnarray}
where
\begin{equation}
\overline {W_{i}¥}=\int W_{1}¥(\theta,\phi)\sin\frac{\theta}{2}
d\theta d\phi,~~~~~i=1,2.
\label{e30}
\end{equation}
Integration over energies is easily performed.  Let us do it for one
particular term in this expression.
\begin{eqnarray}
\int \int &d\varepsilon'_{1}¥&d\varepsilon'_{2}¥
(1-n_{0}¥^{+}¥(\varepsilon'_{1}¥)) n_{0}¥^{+}¥(\varepsilon'_{1}¥+
\varepsilon'_{2}-¥\varepsilon_{1}¥)(1-n_{0}¥^{+}¥(\varepsilon'_{2}¥))\nonumber\\
=T^{2}¥\int\int &dx&dy(1-f(x))f(x+y-t+h)(1-f(y))=
T^{2}¥\int dxf(-x)\frac{x+h-t}{e^{x+h-t}¥-1}\nonumber\\
&=&T^{2}¥\frac{\pi^{2}¥+(h-t)^{2}¥}{2} f(h-t).
\label{e31}
\end{eqnarray}
Here $x=(\varepsilon'_{1}¥-\mu-\gamma H/2)/T$, $y=(\varepsilon'_{2}¥-\mu-\gamma H/2)/T$,
$h=\gamma H/2T$, and $t=(\varepsilon-\mu)/T$, $f(x)=(e^{x}¥ +1)^{-1}¥$. 
It should be stressed that the definitions of variables of integration
$x$ and $y$ depend on particular products like $n^{+}¥(1-n^{-}¥)n^{-}¥$
under the integral.  On the contrary, the variable $t$ and the parameter
$h$ have an invariant definition for all the terms.  Thus we obtain
\begin{eqnarray}
\delta &\hat I_{coll}¥& =-\delta \hat n_{\bf
k}¥\frac{m^{*}¥^{3}¥}{2(2\pi)^{5}¥} (2\overline {W_{1}¥}+\overline
{W_{2}¥}) \frac{T^{2}¥}{2}\left
[(\pi^{2}¥+(h-t)^{2}¥)f(h-t)\right.\nonumber\\
&+&
\left.(\pi^{2}¥+(h+t)^{2}¥)f(-h-t)+(\pi^{2}¥+(t-h)^{2}¥)f(t-h)+
(\pi^{2}¥+(t+h)^{2}¥)f(t+h)\right]\nonumber\\
&=&-\delta \hat n_{\bf
k}¥\frac{m^{*}¥^{3}¥}{2(2\pi)^{5}¥} (2\overline {W_{1}¥}+\overline
{W_{2}¥})\left[(\pi T)^{2}¥+\left(\frac{\gamma H}{2}\right )^{2}¥
+(\varepsilon-\mu)^{2}¥\right]
\label{e32}
\end{eqnarray}
Now we must substitute this expression in the right-hand side of
equation (\ref{e25})
\begin{eqnarray}
\frac{1}{2}(1&+&\frac{F_{1}¥^{a}¥}{3})\frac{1}{2}Sp\int d \tau v_{Fi}¥
(\mbox{\boldmath$\hat \sigma $}\delta \hat I_{coll}¥)\nonumber\\
&=&
-\frac{1}{2}(1+\frac{F_{1}¥^{a}¥}{3})\frac{m^{*}¥^{3}¥}{2(2\pi)^{5}¥} (2\overline {W_{1}¥}+\overline
{W_{2}¥})\int d \tau v_{Fi}¥\left[(\pi T)^{2}¥+\left(\frac{\gamma B}{2}\right )^{2}¥
+(\varepsilon-\mu)^{2}¥\right] \delta\mbox{\boldmath $\sigma $}_{\bf
k}¥({\bf r},t)
\label{e33}
\end{eqnarray}
Taking into account the definition (\ref{e17}) one can directly express
part of this integral containing $(\pi T)^{2}¥+(\gamma H/2)^{2}¥$ through
the spin current density.  The integral as a whole is not in general
expressed in terms of the current. That prevents to consider equations
(\ref{e24}), (\ref{e25}) as the closed system of equations for the
spin density ${\bf M}$ and the spin current density ${\bf J}_{i}¥$. 
However, one can make an assumption which is plausible for weakly
polarized Fermi liquid that the energy dependence of
$\delta\mbox{\boldmath $\sigma $}_{\bf k}¥({\bf r},t)$ is factorized
from the space and direction of ${\bf \hat k}$ dependences:
\begin{equation}
\delta\mbox{\boldmath $\sigma $}_{\bf k}¥({\bf r},t)~
\propto~(n_{0}¥^{+}¥(\varepsilon)-n_{0}¥^{-}¥(\varepsilon))
({\bf A}({\bf r},t)+{\bf B}_{i}¥({\bf r},t)\hat k_{i}¥)
\label{e34}
\end{equation}
In this case in the lowest order of the ratio $\gamma H/\mu$ one can rewrite
the expression (\ref{e33}) as
\begin{equation}
-\frac{m^{*}¥^{3}¥}{6(2\pi)^{5}¥} (2\overline {W_{1}¥}+\overline{W_{2}¥})
\left[(2\pi T)^{2}¥+(\gamma H)^{2}¥\right]{\bf J}_{i}¥({\bf r},t)
\label{e35}
\end{equation}

\section{Results and Discussion}

So, finally we have come to the closed system of equations for the
spin density ${\bf M}$ and the spin current density ${\bf J}_{i}¥$,
\begin{equation}
\frac{\partial {\bf M}}{\partial t} +\frac{\partial 
{\bf J}_{i}¥}{\partial x_{i}¥} -{\bf M}\times \gamma {\bf H}_{0}¥({\bf r})
-({\bf M}^{\parallel}¥+\delta{\bf M}^{\parallel}¥({\bf r}))\times \gamma
{\bf h}({\bf r}, t) =0,
\label{e36}
\end{equation}
\begin{equation}
\frac{\partial {\bf J}_{i}¥}{\partial t} +
\frac{1}{3}v_{F}¥^{2}¥(1+F_{0}¥^{a}¥)(1+\frac{F_{1}¥^{a}¥}{3})
\frac{\partial{\bf M}}{\partial x_{i}¥} - {\bf J}_{i}¥\times\gamma{\bf
H}_{0}¥({\bf r})
+\frac{4}{N_{0}¥}(F_{0}¥^{a}¥-\frac{F_{1}¥^{a}¥}{3})({\bf J}_{i}¥
\times{\bf M^{\parallel}¥})=-\frac{{\bf J}_{i}¥}{\tau},
\label{e37}
\end{equation}
where the current relaxation time is
\begin{equation}
\frac{1}{\tau}=\frac{m^{*}¥^{3}¥}{6(2\pi)^{5}¥} (2\overline {W_{1}¥}+\overline{W_{2}¥})
\left[(2\pi T)^{2}¥+(\gamma H)^{2}¥\right].
\label{e38}
\end{equation}
This system has the same structure as the Leggett spin dynamic equations
\cite{16}.  However, unlike the Leggett equations these are linear
equations for the ${\bf M}$ and ${\bf J}_{i}¥$ in presence of small
but finite polarization.  They contain all the information about spin
rotation or Leggett-Rice effect.  The current relaxation time is
proved to be finite at zero temperature thereby the result that has been
obtained earlier for the dilute Fermi gas  \cite{9,10} is confirmed. 
 
The damping response of the transverse magnetization on the transverse
rf field ${\bf h}({\bf r}, t)$ has been found in \cite{23} and in more
rigorous manner in \cite {22}.  The corresponding transverse
magnetization fluctuations can be established by means of standard 
fluctuation-dissipation relations.

The known dispersion law of the transversal spin waves following from 
equations (\ref{e36}), (\ref{e37}) is
\begin{equation}
\omega=\omega_{L}¥+ (D^{\prime\prime}¥-iD^{\prime}¥)k^{2}¥,
\label{e39}
\end{equation}
where $\omega_{L}¥=\gamma H_{0}¥$ is the Larmor frequency,
\begin{equation} 
D^{\prime}¥=\frac{v_{F}¥^{2}¥(1+F_{0}¥^{a}¥)(1+F_{1}¥^{a}¥/3)\tau}
{3(1+(\kappa \tau\gamma H)^{2}¥)},~~~~~~D^{\prime\prime}¥= \kappa
\tau\gamma H D^{\prime}¥
\label{e40}
\end{equation}
are correspondingly the diffusion coefficient and its reactive part,
$\kappa=F_{0}¥^{a}¥-F_{1}¥^{a}¥/3$.
The spin wave damping is determined by diffusion 
coefficient $D^{\prime}¥$.
In hydrodynamic region $|\kappa| \tau\gamma H\ll 1$ its temperature dependence
is determined by the time of scattering $\tau \propto T^{-2}¥$.
Then, passing through the maximum at $|\kappa| \tau\gamma H \sim 1$
(Leggett-Rice effect, see \cite{26}), at lower temperatures $|\kappa|
\tau\gamma H > 1$ the diffusion coefficient starts to be proportional
to the number of collisions between the quasiparticles $\propto
\tau^{-1}¥$.  Finally, at very low temperatures $T< \gamma H/2\pi$ and
finite polarization it comes to the finite constant value .  Thus the
transverse spin waves in a Fermi liquid with finite polarization have
a finite attenuation at T=0.

The behavior of transverse spin wave damping is similar to, and in fact
has the same origine as, the attenuation of ultrasound with frequency
$\omega$ in a degenerate Fermi liquid \cite{19,2}.  The latter
decreases as $\tau \propto T^{-2}¥$ in hydrodynamic region $\omega \tau \ll
1$, then it passes through the maximum at $\omega \tau \sim 1$ and
behaves as the number of collisions $\propto 1/\tau\propto
(\omega^{2}¥+ (2\pi T)^{2}¥)$ in collisionless region $\omega \tau \gg
1$ (see \cite{27}).  It keeps the finite value $\propto \omega^{2}¥$
at $T=0$ (see \cite{20}).

It will be appropriate to repeat here the
conditions under which our derivation is valid.  The most important is
that we were working in the linear on the space and time variations of
the transverse part of vectorial quasiparticle distribution function. 
In polarized Fermi liquid the large deviations of the magnetization
direction are always accompanied by the changes of its longitudinal
part, therefore we cannot uncouple the kinetic equations for the scalar
and vectorial distribution functions in a Fermi liquid with finite
polarization.  We also lose the possibility to transform the matrix
products in the collision integral as we did.

Thus the important point in our treatment of transversal spin motion was its
independence of longitudinal degrees of freedom.  So long we consider
only linear dynamics there is no coupling between transversal and
longitudinal parts unless we do not take into account  spin nonconserving
collisions (see below).  In that sense we do not see a necessity in
limitations of the developed here theory like $|\omega
-\omega_{L}¥|\tau \ll 1$ (see equation (\ref{e39})) with both $\tau
^{\perp}¥ $ given by (\ref{e38}) and $\tau^{\parallel}¥\propto
T^{-2}¥$ as discussed in \cite{10}.  Nevertheless, it is worth 
noting that till now the measurements have been performed in frame of
these hydrodynamic conditions \cite{28}.

Another important
assumption is the condition of the weak polarization $\gamma
H\ll\varepsilon_{F}¥$ of degenerate Fermi liquid with
$T\ll\varepsilon_{F}¥$, however, with arbitrary relation between
polarization $\gamma H$ and temperature $T$.  This confines all the
quasiparticle momenta to lie in the vicinity of the Fermi surface of
nonpolarized Fermi liquid, and therefore one may decouple the energy
and the angular variables in the collision integral like it was done
in the article \cite{2} and finally to obtain the closed system of the
equations.

Experimentally, the transverse spin current relaxation time is
determined by measurements of the spin echo attenuation or damping of the
standing spin waves.  The results on whether or not the transverse
relaxation time saturates at low temperatures in spin-polarized Fermi
liquids so far have been contradictory.  The recent spin echo
experiments the most probably suggest that spin current relaxation
remains finite as temperature tends to zero \cite{29,30,31}.  At first
sight the large angle deviations, which are the principal feature of
the spin echo method, prevent of application of our theory where the
transverse spin attenuation is calculated in linear on the transverse
perturbations approximation.  Nevertheless, the linear theory does
work because the fastly varying in time "transverse" magnetization is
always kept small during the whole course of the spin echo experiment
if one choose as the natural axis of quantization slowly varying in
time the direction of local magnetization \cite{32}.

On the contrary, the direct measurements of spin waves \cite{33}
demonstrates  much smaller damping than expected on the basis of the
spin echo experiments.  Here, however, the absence of zero-temperature
attenuation is probably masked by not enough precise temperature
determination \cite{34}.

\section  {Concluding remarks: spin polarized Fermi liquid versus
ferromagnetic "Fermi liquid"}

The established zero-temperature attenuation of transversal spin
waves in polarized Fermi liquid is transferred to longitudinal modes like
paramagnons and sound waves via spin nonconserving magnetic
dipole-dipole interaction \cite{35}.  In its turn the finite damping
of longitudinal collective motions causes the finite damping of the
Fermi-liquid quasiparticles.  The latter means that, strictly speaking, a
polarized Fermi liquid at $T=0$ is not the Fermi liquid.  This effect,
however, being proportional to the square of the amplitude of
dipole-dipole interaction will manifest itself at extremely low
temperatures.

A Fermi-liquid theory for the spin waves in a ferromagnetic metal has been
developed by Abrikosov and Dzialoshinskii \cite{36}.  It was done in
neglecting of quasiparticle collisions, hence the dampingless spectrum
has been obtained.  However, the following was noted by Herring \cite
{37}: "For a ferromagnetic metal\ldots.  if the spin of quasiparticle
at the Fermi surface is reversed, the corresponding quasiparticle
state will no longer be closed to the Fermi surface, and it will have a
finite, rather than an inifinitesimal, decay rate."  The finite decay
rate of quasiparticle states produces the zero-temperature spin wave
attenuation.  The latter certainly contradicts to the Goldstone
theorem for a isotropic ferromagnetic ground state.  So, starting from a
Fermi-liquid approach to the itinerant ferromagnet we come to the
contradiction.  The resolution of this paradox is that in an itinerant
ferromagnet the formation of off-diagonal deviations of the 
momentum-dependent 
disribution function $\delta \mbox{\boldmath $\sigma $}_{\bf
k}¥({\bf r},t) $ and its time-space variations according to the
kinetic equation will be blocked up by the alteration of orbital part
of the electron wave function and corresponding increase of an
interaction energy.  So, the equation of motion of magnetic degrees of
freedom is not formulated in $({\bf k},{\bf r})$ or phase space but
only in the space of coordinates ${\bf r}$ as has been done by
Landau and Lifshits \cite{38} for the magnetization density 
$ {\bf M}({\bf r},t)$.

\section{Acknowledgments}

I am grateful to I.A.Fomin who had stimulated my
interest in the problem.  I express my acknowledgment to G.A.Vermeulen for
valuable help and numerous wholesome discussions.  Also I am indebted
to D.I.Golosov for very interesting conversations on the subject as well to 
E.Kats, T.Ziman, G.Jackeli for valuable remarks and questions.
I would like to thank O.Buu and W.Mullin for the interest in my results and
explanations concerning spin echo, and D.Candela and M.Meizel for
sending me their recent papers.


\end{document}